\documentclass[aps,prd,twocolumn,eqsecnum,showpacs,amsmath]{revtex4}

\usepackage{subfigure}
\usepackage[dvips]{color,graphicx}
\usepackage{amsfonts,amssymb,theorem}
\newcommand{\D}{{\rm d}}

{\theorembodyfont{\upshape}
}
{\theorembodyfont{\upshape}
}
{\theorembodyfont{\upshape}
}
{\theorembodyfont{\upshape}
}
{\theorembodyfont{\upshape}
}
{\theorembodyfont{\upshape}
}

\newcommand{\dalm}{\kern1pt\vbox{\hrule height 0.9pt\hbox{\vrule width 0.9pt\hskip 2.5pt\vbox{\vskip 5.5pt}\hskip 3pt\vrule width 0.3pt}\hrule height 0.3pt}\kern1pt}

\begin{document}


\title{
Cosmological wormholes
}

\author{
$^{1}$Hideki Maeda\footnote{Electronic
address:hideki@cecs.cl},
$^{2}$Tomohiro Harada\footnote{Electronic
address:harada@rikkyo.ac.jp} and 
$^{3,4}$B.~J.~Carr\footnote{Electronic address:B.J.Carr@qmul.ac.uk}}
\affiliation{
$^{1}$Centro de Estudios Cient\'{\i}ficos (CECS), Arturo Prat 514, Valdivia, Chile\\
$^{2}$Department of Physics, Rikkyo University, Toshima, Tokyo 171-8501, Japan\\
$^{3}$Astronomy Unit, Queen Mary, University of London, Mile End Road, London E1 4NS, UK\\
$^{4}$Research Center for the Early Universe, Graduate School of Science, University of Tokyo, Tokyo 113-0033, Japan}
\date{\today}

\begin{abstract}                
Motivated by the cosmological wormhole solutions obtained from our recent numerical investigations,  
we provide a definition of a wormhole which applies to dynamical situations.
Our numerical solutions do not have timelike trapping horizons but they are wormholes in the sense that they connect two or more asymptotic regions. 
Although the null energy condition must be violated for static wormholes, we find that it can still be satisfied in the dynamical context.
Two analytic solutions for a cosmological wormhole connecting two Friedmann universes without trapping horizons are presented.
\end{abstract}


\maketitle

\section{Introduction} 
A wormhole is a hypothetical object in general relativity which connects two or more asymptotic regions.
In the history of wormhole research, the 
Morris-Thorne solution has occupied a central position as a typical static wormhole~\cite{mt1988}, although many static wormhole metrics were obtained before that~\cite{ellis1973,before}.
Subsequent research revealed that wormholes may admit 
superluminal travel due to the global spacetime topology~\cite{visser,superluminal,lobo2007} and they may also lead to time machines~\cite{mty1988,timemachine}.

It is well known that these intriguing static configurations require the violation of the null energy condition~\cite{visser,hv1997,negative}.
In the asymptotically flat case, this is also a consequence of the
topological censorship~\cite{TC}. On the other hand,
wormhole spacetimes can be constructed with arbitrarily small violation of the averaged null energy condition~\cite{vkd2003}. This
suggests that the wormhole configuration could be realized merely by quantum effects violating the energy conditions.

Dynamical wormholes are not as well understood as static wormholes.
Their comprehensive study was pioneered by Hochberg and Visser~\cite{hv1998} and Hayward~\cite{hayward1999}, who introduced two independent 
quasi-local definitions of a wormhole throat in a dynamical spacetime.
In these definitions, wormhole throats are trapping horizons~\cite{hayward1994} of various kinds, but the null energy condition must still be violated there.

Recently, 
we have numerically found an interesting one-parameter family of spherically symmetric dynamical wormhole solutions in an accelerating Friedmann background~\cite{mhc1,hmc1}.
The spacetime contains a perfect fluid and admits a homothetic Killing vector. This requires the equation of state to be linear and the cosmic expansion is accelerating for an appropriate equation of state parameter.
The wormhole solutions are asymptotically Friedmann at one infinity and they have another infinity, which may also be asymptotically Friedmann 
for a special value of the parameter which describes these solutions. In this case, the wormhole throat connects two Friedmann universes.
Interestingly, in this class of dynamical wormhole spacetimes, the dominant energy condition is satisfied everywhere. In fact, the Hochberg-Visser and Hayward conditions are avoided because the spacetimes are trapped everywhere and there is no trapping horizon.

In the Hochberg-Visser and Hayward definitions, a wormhole throat is a two-dimensional surface of nonvanishing minimal area on a null hypersurface.
However, there is no past null infinity in our dynamical wormhole solutions because there exists an initial singularity.
The wormhole throats are therefore not defined on a null hypersurface but on a spacelike hypersurface. 
This demonstrates that the Hochberg-Visser and Hayward definitions miss an important class of dynamical wormhole spacetimes, namely {\it cosmological wormholes} which are asymptotically Friedmann universe and start with a big-bang singularity.

In this paper, we define a wormhole quasi-locally in terms of a surface of nonvanishing minimal area on a {\it spacelike} hypersurface and compare its properties with those of the Hochberg-Visser or Hayward wormhole. 
We also construct two analytic examples of
cosmological wormholes.
One corresponds to the numerical solution 
obtained in refs.~\cite{mhc1,hmc1}, but it contains a singular hypersurface which 
violates the null energy condition.
The other is a smooth model involving a combination of a perfect fluid and 
a ghost scalar field but the total matter content still satisfies the dominant 
energy condition.

The plan of this paper is as follows.
In section~II, basic equations are given. 
In section~III, we present several possible quasi-local definitions of the wormhole throat for the static and dynamical cases and discuss the violation of the energy conditions.
In section~IV, we construct two analytic 
cosmological wormholes, one of which contains massive thin shells.
Section~V makes concluding remarks and discusses future prospects.

\section{Formulation}
For simplicity, we assume spherical 
symmetry throughout this paper. The metric signature convention is taken to be 
$(-,+,+,+)$, with
Greek indices running over spacetime coordinates.
We follow the notation of Hayward~\cite{hayward1996}, in which the line element 
is written locally in double-null coordinates as
\begin{eqnarray}
ds^2 = -2e^{-f}d\xi_{-}d\xi_{+}+r^2 d\Omega^{2},
\end{eqnarray}  
where $d\Omega^{2}:=d\theta^{2}+\sin^{2}\theta d\phi^{2}$ and $f=f(\xi_{-},\xi_{+})$.
The function $r=r(\xi_{-},\xi_{+})$ is defined such that the area of the metric sphere is $4\pi r^2$ and we put $c=1$. 
The spacetime will be assumed time-orientable, 
with $\partial /\partial \xi_{+}$ and $\partial /\partial \xi_{-}$ 
being future-pointing. 
We denote  $\partial /\partial \xi_{+}$ and $\partial /\partial \xi_{-}$ 
by $\partial_{+}$ and $\partial_{-}$, 
these being the outgoing and ingoing null normal 
vectors, respectively.

We use $A$ and $B$ as indices on the two-dimensional spacetime 
spanned by $\partial_{+}$ and $\partial_{-}$ and $|$ as the associated covariant 
derivative.
The Misner-Sharp mass $m$~\cite{ms1964} is then given by
\begin{eqnarray}
\label{qlm2}
m := \frac{r}{2G}(1-r^{|A}r_{|A})
=\frac{r}{2G}\left(1+\frac12 r^2e^{f}\theta_{+}\theta_{-}\right),
\end{eqnarray}  
where the area expansions 
along $\partial_{+}$ and $\partial_{-}$ are defined respectively as
\begin{eqnarray}
\theta_{+}&:=2r^{-1}\partial_{+}r,\\
\theta_{-}&:=2r^{-1}\partial_{-}r.
\end{eqnarray}  
We assume $\theta_{+}\geq \theta_{-}$, at least locally, 
as is always possible.

The tangent vectors of the radial null geodesics $l^\mu_{\pm}$ are given by 
\begin{eqnarray}
l^\mu_{+}\frac{\partial}{\partial x^\mu}&:=&h_{+}e^f\frac{\partial}{\partial \xi_{+}},\\
l^\mu_{-}\frac{\partial}{\partial x^\mu}&:=&h_{-}e^f\frac{\partial}{\partial \xi_{-}},
\end{eqnarray}  
where $h_{+}=h_{+}(\xi_{-})$ and $h_{-}=h_{-}(\xi_{+})$ are functions of $\xi_-$ and $\xi_+$, respectively, and $h_{\pm}$ must be positive for $l^\mu_{\pm}$ to be future-pointing.
The expansions of the null geodesics are given by
\begin{eqnarray}
\Theta_{\pm}:=l^\mu_{\pm;\mu}=h_{\pm} e^{f} \theta_{\pm},
\end{eqnarray}  
where the semicolon denotes the covariant derivative 
associated with the four-dimensional metric $g_{\mu\nu}$.

The notion of trapping horizons was introduced by Hayward to give a quasi-local definition of black holes~\cite{hayward1994,hayward1996}. 
A metric sphere is said to be {\it trapped} if $\theta_{+}\theta_{-}>0$,
{\it untrapped} if $\theta_{+}\theta_{-}<0$,
and {\it marginal}  if $\theta_{+}\theta_{-}=0$.
If $e^{f}\theta_{+}\theta_{-}$ 
has non-vanishing derivative,
the spacetime is divided into trapped and untrapped regions, 
separated by marginal hypersurfaces.
A marginal sphere is said to be {\it future} if $\theta_{+}=0$, {\it past} if
$\theta_{-}=0$, and {\it bifurcating} if $\theta_{\pm}=0$.
A future marginal sphere is 
{\it outer} if
$\partial_{-}\theta_{+}<0$, {\it inner} if $\partial_{-}\theta_{+}>0$, 
and {\it degenerate} if $\partial_{-}\theta_{+}=0$.
A past marginal sphere is 
{\it outer} if $\partial_{+}\theta_{-}<0$, {\it inner} if $\partial_{+}\theta_{-}>0$, 
and {\it degenerate} if $\partial_{+}\theta_{-}=0$.
A {\it trapping horizon} is the closure of a hypersurface foliated by future or past, outer or inner
marginal spheres~\cite{hayward1994,hayward1996}.
From the definition of the Misner-Sharp mass, 
one can easily show that $r=2Gm$ on trapping horizons. 
A {\it future (past) outer trapping horizon} is 
the closure of a hypersurface foliated by future (past) outer marginal 
spheres and this is the counterpart of a black hole (white hole) 
apparent horizon.
Accordingly, we call the closure of a hypersurface foliated by 
bifurcating marginal spheres a {\it bifurcating trapping horizon}. 

Here we show that above definitions are equivalent even if we replace
$\theta_{\pm}$ and $\partial_{\pm}$ with $\Theta_{\pm}$ and
$d/d\lambda_{\pm}$, respectively, where $\lambda_{\pm}$ are the 
affine parameters of the null geodesics.
We obtain 
\begin{align}
\frac{d\Theta_{\pm}}{d\lambda_{\pm}}&=(\theta_{\pm}\partial_{\pm}f+
\partial_{\pm}\theta_{\pm})h_{\pm}^2e^{2f},\\
\frac{d\Theta_{\pm}}{d\lambda_{\mp}}&=(\theta_{\pm}
\partial_{\mp}h_{\pm}+h_{\pm}\theta_{\pm}\partial_{\mp}f+
h_{\pm}\partial_{\mp}\theta_{\pm})h_{\pm}e^{2f}.
\end{align}
Thus we have 
\begin{eqnarray}
\frac{d\Theta_{\pm}}{d\lambda_{\pm}}&=&h_{\pm}^2e^{2f}\partial_{\pm}\theta_{\pm},\\
\frac{d\Theta_{\pm}}{d\lambda_{\mp}}&=&h_{\pm}^2e^{2f}\partial_{\mp}\theta_{\pm}
\end{eqnarray}
on the hypersurface with $\theta_{\pm}=0$, where we assume that the metric and $h_{\pm}$ are at least $C^1$.
It is therefore clear that the signs of $\Theta_{\pm}$,
$d\Theta_{\pm}/d\lambda_{\pm}$, and $d\Theta_{\pm}/d\lambda_{\mp}$ are
the same as those of $\theta_{\pm}$, 
$\partial_{\pm} \theta_{\pm}$, and $\partial_{\mp} \theta_{\pm}$, respectively.

The most general stress-energy tensor under spherical symmetry is given by
\begin{eqnarray}
T_{\mu\nu}dx^\mu dx^\nu &
=T_{--}d\xi_{-}^2+2T_{-+}d\xi_{-}d\xi_{+} \nonumber \\
&+T_{++}d\xi_{+}^2+p r^2 d\Omega^2,
\end{eqnarray}  
where $T_{--}$, $T_{-+}$, $T_{++}$, and $p$ are functions of $\xi_{-}$ and $\xi_{+}$.
The Einstein equations then become
\begin{eqnarray}
&&\partial_{-}\partial_{-}r+\partial_{-}f\partial_{-}r=-4\pi Gr T_{--}, \label{nullbasic1} \\
&&\partial_{+}\partial_{+}r+\partial_{+}f\partial_{+}r=-4\pi Gr T_{++}, \label{nullbasic2}\\
&&r\partial_{+}\partial_{-}r+\partial_{-}r\partial_{+}r+\frac{1}{2}e^{-f}=4\pi Gr^2T_{-+}, \label{nullbasic3}\\
&&r^2 \partial_{+}\partial_{-}f+2\partial_{-}r\partial_{+}r
+e^{-f} \nonumber \\
&&~~~~~~~~~~~~~~~~~~~~~=8\pi G r^2(T_{-+}+e^{-f}p). \label{nullbasic4}
\end{eqnarray}  
The null energy condition for the matter field implies
\begin{eqnarray}
T_{--}\ge 0,~~~T_{++} \ge 0,
\end{eqnarray}
while the dominant energy condition implies
\begin{eqnarray}
T_{--} \ge 0,~~T_{++}\ge 0,~~T_{-+}\ge 0.
\end{eqnarray}  
The dominant energy condition assures that a causal observer measures 
non-negative energy density and that the energy flux is a future-directed causal vector.
The dominant energy condition implies the null energy condition.

\section{Definitions of wormholes}
In this section, we discuss several definitions of 
dynamical wormholes.
Although the concept of a wormhole is originally topological 
and global, it is possible to define it quasi-locally in terms of 
two-dimensional spheres of minimal area.
First we revisit static wormholes and then consider the possible generalization to the dynamical case.

\subsection{Static wormholes}

Staticity 
is defined by the existence of a 
timelike Killing vector $\partial_{t}$. We can write this as
$\partial_{t}=(\partial_{-}+\partial_{+})/\sqrt{2}$, where  
$t=(\xi_{-}+\xi_{+})/\sqrt{2}$ is the Killing time coordinate.
Then we have
\begin{eqnarray}
\partial_{t}r=(\partial_{-}r+\partial_{+}r)/\sqrt{2}=0,\label{r-static}\\
\partial_{t}f=(\partial_{-}f+\partial_{+}f)/\sqrt{2}=0.
\end{eqnarray}
There also exists 
another natural coordinate
$x=(-\xi_{-}+\xi_{+})/\sqrt{2}$,
corresponding to $\partial_{x}=(-\partial_{-}+\partial_{+})/\sqrt{2}$.
We define a {\it static wormhole} as a timelike hypersurface foliated by 
minimal spheres on the constant $t$ spacelike hypersurfaces.
At a minimal sphere, we have
\begin{eqnarray}
\partial_{x}r=(-\partial_{-}r+\partial_{+}r)/\sqrt{2}=0. 
\label{throat-static}
\end{eqnarray}  
Then, from Eq. (\ref{throat-static}) with $\theta_+ \ge \theta_-$,
we obtain $\theta_{+}=\theta_{-}=0$ there. We conclude that
a wormhole throat in static spacetimes is a timelike 
bifurcating trapping horizon.

Differentiating Eq.~(\ref{r-static}) with respect to $\xi_{-}$ 
and $\xi_{+}$, we obtain  
\begin{eqnarray}
\partial_{-}\partial_{-}r=\partial_{+}\partial_{+}r=-\partial_{-}\partial_{+}r. \label{ruu=-rvv}
\end{eqnarray}
Equations~(\ref{nullbasic1}) and (\ref{nullbasic2}), 
together with the condition $\theta_{+}=\theta_{-}=0$, imply 
\begin{eqnarray}
\partial_{-}\partial_{-}r=-4\pi Gr T_{--} \label{ruu}
\end{eqnarray}  
and 
\begin{eqnarray}
\partial_{+}\partial_{+}r=-4\pi Gr T_{++} \label{rvv}
\end{eqnarray}  
on the wormhole throat.
The minimality of the area then yields
\begin{equation}
\partial_{x}\partial_{x}r>0.
\label{rxx}
\end{equation}
Equations~(\ref{ruu=-rvv}) and (\ref{rxx}) imply
$\partial_{-}\partial_{-}r>0$ and $\partial_{+}\partial_{+}r>0$,
so $T_{--}<0$ and $T_{++}<0$ at the wormhole throat from Eqs.~(\ref{ruu}) and (\ref{rvv}). 
Therefore the null energy condition is 
violated at a wormhole throat in a static spacetime~\cite{visser}.

It should be noted that all the definitions of dynamical wormholes 
will reduce to the standard one
in the static case. 
In this sense, we are generalizing the notion of 
a static wormhole to the dynamical situation.

\subsection{Wormholes as minimal spheres on null hypersurfaces}
Hochberg and Visser~\cite{hv1998} and Hayward~\cite{hayward1999} 
define a wormhole throat in terms of null expansions but in slightly different ways, which we now discuss.

In the spherically symmetric case, Hochberg and Visser~\cite{hv1998} define a wormhole throat by $\theta_{+}=0$ and $\partial_{+}\theta_{+} > 0$ or $\theta_{-}=0$ and $\partial_{-}\theta_{-}>0$.
Although the original definition allowed equality, we do not consider that case here.
We thereby avoid the Killing horizon in Schwarzschild spacetime being a wormhole throat.
On the other hand, a Hayward traversable wormhole throat is a
{\it timelike} outer trapping horizon, i.e., a timelike 
trapping horizon with $\theta_{+}=0$ and $\partial_{-}\theta_{+}<0$
or $\theta_{-}=0$ and $\partial_{+}\theta_{-}>0$.

Let $\eta^{A}$ be the timelike generator of the Hayward wormhole throat with $\theta_{+}=0$. Then 
$\zeta^{+}\partial_{+}\theta_{+}+\zeta^{-}\partial_{-}\theta_{+}=0$
and, using $\zeta^{+}\zeta^{-}<0$ 
and $\partial_{-}\theta_{+}>0$, we find $\partial_{+}\theta_{+}>0$.
Similarly, $\partial_{-}\theta_{-}>0$ holds for the Hayward wormhole 
throat with $\theta_{-}=0$.
Therefore, a Hayward wormhole throat is necessarily 
a Hochberg-Visser wormhole throat but the opposite does not hold in general.
In fact, the Hochberg-Visser wormhole throat may be spacelike.
In both cases, $\partial_{+}\theta_{+} > 0$ or $\partial_{-}\theta_{-}>0$ is satisfied on the wormhole throat, which means that it is a minimal sphere on null hypersurfaces.

With the Hochberg-Visser or Hayward definition, 
the infinitesimal sectional area of the null geodesic congruence 
reaches a minimum at the wormhole throat. 
Also both definitions of a wormhole throat are based
on the physical intuition obtained from the asymptotically flat examples.
However, in the cosmological situations this intuition may not apply 
because there is an 
initial singularity. 

We now show that the null energy condition is violated 
on the Hochberg-Visser and Hayward wormhole throats.
Although we consider the $\theta_+=0$ case below, the argument also 
applies for $\theta_{-}=0$.
From Eqs.~(\ref{nullbasic2}) and (\ref{nullbasic3}), we obtain
\begin{eqnarray}
\partial_{+}\theta_{+}&=&-8\pi GT_{++}, \label{mm3}\\
\partial_{-}\theta_{+}&=&8\pi GT_{-+}-\frac{1}{r^2}e^{-f} \label{mm4}
\end{eqnarray}  
on the trapping horizon with $\theta_{+}=0$. 
Eq.~(\ref{mm3}) then immediately implies the violation of 
the null energy condition for the Hochberg-Visser throat.
Since we have shown that a Hayward throat is necessarily a Hochberg-Visser throat, the null energy condition is violated also for the Hayward wormhole throat.

It is important to 
investigate whether wormholes always 
need exotic matter fields
which violate some energy condition.
In these generalizations~\cite{hv1998,hayward1999},
the null energy condition must be violated at the wormhole throat, so
the existence of traversable wormholes might seem implausible.
However, as we will 
see, these definitions 
are not well-motivated in a cosmological background.

\subsection{Wormholes as minimal spheres on spacelike hypersurfaces}
We now present a definition of wormholes 
for spherically symmetric spacetimes which 
is relevant to the problems 
raised above. 
To accomplish this, we consider a spherically symmetric 
spacelike hypersurface and define 
a minimal sphere with $r>0$ on this hypersurface. 
This means 
\begin{eqnarray}
r_{|A}\zeta^{A}&=&0, \label{throat1} \\
r_{|AB}\zeta^{A}\zeta^{B} &>& 0, \label{throat2}
\end{eqnarray}  
where 
$\zeta^{A}$ is any 
nonvanishing radial spacelike vector (i.e. one with 
$g_{AB}\zeta^{A}\zeta^{B}>0$).
It should be noted
that any definition in terms of time-slicing will inevitably 
entail the problem of slice-dependence.

Here we note that $(r_{|A}\zeta^A)_{|B}\zeta^B>0$ could be an
alternative definition to Eq.~(\ref{throat2}) because the left-hand side
gives the second order derivative along $\zeta^{A}$.
It should be noted, however, that 
the definition (\ref{throat2}) does not involve 
the derivative of $\zeta^A$.
Actually, $(r_{|A}\zeta^A)_{|B}\zeta^B>0$ is equivalent to 
Eq.~(\ref{throat2}) if $\zeta^{A}$ is tangent to a geodesic.

We say that a timelike hypersurface is 
a wormhole throat if it is 
foliated by minimal spheres 
on a spacelike hypersurface of the time-slicing.
This 
reduces to the definition of a static wormhole for static spacetimes
if we take the constant Killing time hypersurface.
For general dynamical spacetimes,
Eq.~(\ref{throat1}) gives either
\begin{eqnarray}
\frac{\zeta^{-}}{\zeta^{+}}=-\frac{\theta_{+}}{\theta_{-}} \label{sign2}
\end{eqnarray}  
or $\theta_{-}=\theta_{+}=0$.
Eq.~(\ref{sign2}) implies $\theta_+ \theta_->0$ 
because $\zeta^{-}\zeta^{+}<0$.
Hence we conclude that a wormhole throat 
is either a trapped sphere or a bifurcating trapping horizon.
This means that 
there are two classes of wormhole throats.
The first is ``locally and momentarily static'' in the sense that it is
a bifurcating trapping horizon.
The second excludes a bifurcating trapping horizon and 
hence cannot be a static wormhole.
By the mean value theorem, there then exists at least one trapping horizon 
if the spacetime admits a wormhole throat and 
if $\theta_{+}\theta_{-}<0$ holds at spacelike infinity
(as in the asymptotically flat spacetime).

Although both Hayward's and Hochberg and Visser's wormholes have a minimal sphere on 
the null hypersurface, we can make this hypersurface 
spacelike by an infinitesimally small deformation.
Therefore if either of
their wormhole exists,
then so does ours, i.e.,  
our wormhole definition generalizes theirs.

It is interesting to examine the implication of 
our wormhole
definition for the energy conditions.
One can easily show that the null energy condition is 
violated on our wormhole throat being a bifurcating trapping horizon. 
The proof 
is essentially the same 
as for the static case.
On the other hand, there may be no violation of 
the energy conditions on a wormhole 
which is a trapped sphere.
This means that wormhole throats defined
in terms of spacelike hypersurfaces would be much more
plausible than those defined in terms of null hypersurfaces.

Here we should comment on the traversability of our wormhole solutions.
Hayward defined a wormhole throat as a temporal (or timelike) trapping
horizon. By contrast, our wormhole definition does not depend on the
existence of a temporal trapping horizon. For example, the maximally
extended Schwarzschild spacetime does not contain a wormhole throat in the
sense of Hochberg-Visser or Hayward but it does contain one in our sense
and it is located inside the horizon~\cite{hayward}. Intuitively, Hayward's
definition of a wormhole throat enables an observer to travel from one
infinity to the other in an asymptotically flat spacetime. In a
cosmological spacetime, however, there is generally no past infinity
because of the initial singularity, so Hayward's concept of a traversable
wormhole might not be suitable. That is why we have adopted an alternative
wormhole definition.

\section{Analytic solutions for the cosmological wormhole}
In the last section, we have introduced a new 
quasi-local definition of a wormhole throat on a spacelike 
hypersurface
and shown that such a throat must coincide with 
a bifurcating trapping horizon or be located in the trapped region.
This implies that 
one can have a wormhole throat in the latter case 
even if there is no trapping horizon in the spacetime.

Recently, we have found an interesting one-parameter family of numerical spherically symmetric dynamical wormhole solutions of this kind~\cite{mhc1,hmc1}.
In this section, we review these
and give two analytic solutions for a wormhole connecting two different Friedmann universes, one of which contains a massive thin shell.

\subsection{Friedmann cosmological wormhole}
Our new class of wormhole solutions are ``cosmological'' in the sense that one of the asymptotic regions is Friedmann~\cite{mhc1}. 
They are spherically symmetric and self-similar and contain a perfect fluid with equation of state of the form $p=(\gamma -1) \mu$ with $0<\gamma<2/3$ (dubbed ``dark energy''). 
This matter model violates the strong energy condition but still satisfies the dominant energy condition.

\begin{center}
\begin{figure}[htbp]
\includegraphics[width=0.5\textwidth]{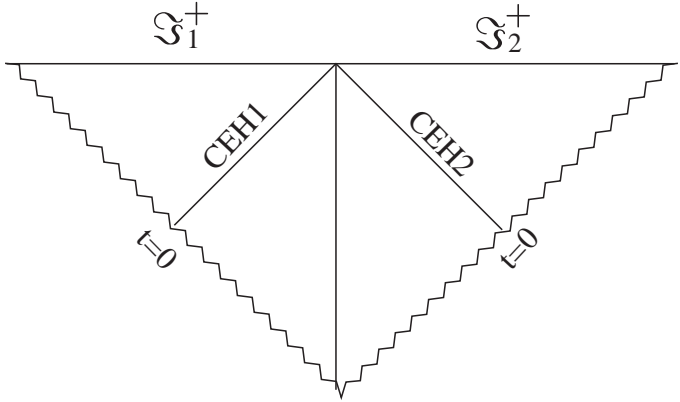}
\caption{\label{fg:F_QF} The causal structure of 
the self-similar solution with dark energy 
which is asymptotically Friedmann at one end 
and (quasi-)Friedmann at the other.
$t=0$ corresponds to the initial singularity, while 
$\Im^{+}_1$ and $\Im^{+}_2$ correspond to two distinct null infinities.
CEH stands for cosmological event horizon.
The spacetime is asymptotically quasi-Kantowski-Sachs spacetime around the wormhole throat~\cite{hmc1}.
}
\end{figure}
\end{center}

These solutions are a subset of the complete family of asymptotically 
Friedmann spherically symmetric self-similar solutions with dark energy 
and are obtained by using the formulation and asymptotic analyses 
presented in ref.~\cite{hmc1}.  
The most interesting wormhole solutions are of two types. 
The first connects two Friedmann universes; the second connects a 
Friedmann universe and a quasi-Friedmann universe (in the sense that 
there is an angle deficit at large distances).
The causal structure of these solutions is shown in Fig.~\ref{fg:F_QF},
which shows that the world-tube of the throat is timelike.
Although the solutions are spherically symmetric and self-similar, 
they are presumably a subset of more general non-self-similar 
spherically symmetric wormhole solutions.
There are no trapping horizons 
but the whole of the spacetime is trapped. Due to the  
expansion of the wormhole throat, it is not
a marginal sphere but a past trapped sphere.

The most intriguing of our numerical cosmological wormhole solutions 
is the one which connects two Friedmann universes.
This could be important as a cosmological model in the very early universe. Refs.~\cite{hmc1,mhc1} give further details.

\subsection{An analytic solution with thin shell}
Here we construct an analytic solution for the Friedmann-Friedmann cosmological wormhole.
Because the spacetime is asymptotically quasi-Kantowski-Sachs around the wormhole throat, we can match a Friedmann exterior to a Kantowski-Sachs interior at a hypersurface $\Sigma$.
Both sides are solutions of the Einstein field equations for a perfect fluid with
$p=(\gamma-1)\mu$ which satisfies
the dominant energy condition 
($0\le  \gamma \le 2$).
We do not consider the $\gamma=0$ case because the treatment 
will then be significantly different. 
The Friedmann solution with coordinates $x^\mu=(t,x,\theta,\phi)$ is given by
\begin{align}
ds_{+}^2&=-dt^2+\biggl(\frac{t}{t_0}\biggl)^{4/(3\gamma)}(dx^2+x^2d\Omega^2), \label{f} \\
8\pi G\mu&=\frac{4}{3\gamma^2t^2},\\
2Gm&=\frac{4x^3}{9\gamma^2t^2}\biggl(\frac{t}{t_0}\biggl)^{2/\gamma},
\end{align}  
where $t_0$ is a positive constant. The Kantowski-Sachs solution with coordinates ${\bar x}^\mu=({\bar t},{\bar x},\theta,\phi)$ is given by
\begin{align}
ds_{-}^2&=-\frac{(2-3\gamma)(2-\gamma)}{\gamma^2}d{\bar t}^2+C_0^2{\bar t}^{4(1-\gamma)/\gamma}d{\bar x}^2+{\bar t}^2d\Omega^2, \label{ks} \\
8\pi G\mu&=\frac{4(1-\gamma)}{(2-3\gamma)(2-\gamma){\bar t}^2},\\
2Gm&=\frac{4(1-\gamma)^2{\bar t}}{(2-3\gamma)(2-\gamma)},
\end{align}  
where $C_0$ is a positive constant and 
$0<\gamma<2/3$ is required for the spacetime to have Lorentzian signature.
We focus on the expanding regions with $t \ge 0$ and ${\bar t} \ge 0$.

We consider a hypersurface $\Sigma_+$ defined on the Friedmann side by
\begin{eqnarray}
t=\lambda, \quad 
x=C_1 \lambda^{-(2-3\gamma)/(3\gamma)}, \quad \theta=\theta, \quad \phi=\phi, \label{hyp-frw}
\end{eqnarray}  
and a hypersurface $\Sigma_-$ defined on the Kantowski-Sachs side by 
\begin{eqnarray}
{\bar t}=\lambda,\quad
{\bar x}=C_2\lambda^{-(2-3\gamma)/\gamma}, \quad \theta=\theta, \quad \phi=\phi, \label{hyp-ks}
\end{eqnarray}  
where $C_1$ and $C_2$ are positive constants and $\lambda$ is a linking variable.
The induced metrics on $\Sigma_+$ and $\Sigma_-$ are 
\begin{align}
ds_{+(3)}^2&=\frac{(2-3\gamma)^2C_1^2-9\gamma^2t_0^{4/(3\gamma)}}{9\gamma^2t_0^{4/(3\gamma)}}d\lambda^2+\frac{C_1^2}{t_0^{4/(3\gamma)}}\lambda^2d\Omega^2, \\
ds_{-(3)}^2&=\frac{(2-3\gamma)^2C_0^2C_2^2-(2-3\gamma)(2-\gamma)}{\gamma^2}d\lambda^2+\lambda^2d\Omega^2, 
\end{align}  
respectively. 
Identifying $\Sigma_+$ with $\Sigma_-$, we require $ds_{+(3)}^2=ds_{-(3)}^2$, i.e. continuity of the induced metric, which implies
\begin{eqnarray}
C_1^2=t_0^{4/(3\gamma)},\quad 
C_2^2=\frac{27\gamma^2-84\gamma+40}{9(2-3\gamma)^2C_0^2}. \label{C-constants}
\end{eqnarray}  
$C_2^2$ is non-negative only for $\gamma \le (14-2\sqrt{19})/9 \simeq 0.5869  (<2/3)$,
so matching is impossible for $(14-2\sqrt{19})/9<\gamma<2/3$.
Since the induced metric is 
\begin{eqnarray}
ds_\Sigma^2&=&h_{ab}\D y^a \D y^b \nonumber \\
&=&-\frac{4(3\gamma-1)}{9\gamma^2}d\lambda^2+\lambda^2d\Omega^2,
\end{eqnarray}  
the matching hypersurface $\Sigma (:= \Sigma_+ \equiv \Sigma_-)$  and coordinate $\lambda$ are timelike for $\gamma>1/3$, spacelike for $\gamma<1/3$ and null for $\gamma=1/3$.

We now show that the whole 
spacetime is trapped.
In the Kantowski-Sachs region, the condition $r<2Gm$ reduces to $\gamma^2>0$, so it is foliated by trapped surfaces.
On the other hand, in the Friedmann region{\bf ,} $r<2Gm$ implies
\begin{eqnarray}
\biggl(\frac{t}{t_0}\biggl)^{-4/(3\gamma)}<\frac{4x^2}{9\gamma^2t^2} \label{trap-f}.
\end{eqnarray}  
Because of Eq.~(\ref{C-constants}) and the condition $0<\gamma<2/3$,  
this reduces to $x>(3\gamma C_1/2) t^{-(2-3\gamma)/(3\gamma)}$ and hence is satisfied in the exterior Friedmann region defined by $x>C_1 t^{-(2-3\gamma)/(3\gamma)}$.

The Friedmann spacetime is ``exterior'' to $\Sigma_-$ in the sense of having a larger value of ${\bar x}$.
To complete the construction of the Friedmann-Friedmann cosmological wormhole, we attach another Friedmann universe to the Kantowski-Sachs spacetime via another matching surface.
That matching hypersurface is also represented by Eq.~(\ref{hyp-frw}) in the Friedmann spacetime but by 
\begin{align}
{\bar t}=\lambda,\quad
{\bar x}=-C_2\lambda^{-(2-3\gamma)/\gamma}, \quad \theta=\theta, \quad \phi=\phi  \label{hyp-ks2}
\end{align}  
in the Kantowski-Sachs spacetime.
The curve (\ref{hyp-ks2}) is the reflection of the curve (\ref{hyp-ks}) about the line ${\bar x}=0$ in the $(t,{\bar x})$ plane.
At the hypersurface (\ref{hyp-ks2}), the Friedmann spacetime is attached to the ``exterior'' of the hypersurface in the sense of having a smaller value of ${\bar x}$.
As a result, the matter distributions on those two matching hypersurfaces are the same.
The resulting spacetime is trapped everywhere and does not contain a trapping horizon.

In the following subsections, we calculate the jump of the second fundamental form on $\Sigma$, which gives the energy-momentum tensor on $\Sigma$~\cite{Poisson}.
We first consider the case where the shell is timelike or spacelike, and then the case where it is null.

\subsubsection{The timelike or spacelike shell case}
We first consider the case with $\gamma \ne 1/3$.
We define two functions $f(t,x)$ and $g({\bar t},{\bar x})$ by
\begin{eqnarray}
f&:=&-x+C_1t^{-(2-3\gamma)/(3\gamma)},\\
g&:=&-{\bar x}+C_2{\bar t}^{-(2-3\gamma)/\gamma},
\end{eqnarray}  
with $\Sigma$ being described by $f=0$ and $g=0$ in the Friedmann and Kantowski-Sachs regions, respectively. 
The unit 1-forms normal to $\Sigma$ are given by
\begin{align}
n_\mu dx^\mu&=\frac{2-3\gamma}{2\sqrt{|3\gamma-1|}}dt+\frac{3\gamma}{2C_1\sqrt{|3\gamma-1|}}t^{2/(3\gamma)}dx,\\
{\bar n}_\mu d{\bar x}^\mu&=\frac{\sqrt{(2-3\gamma)(2-\gamma)(27\gamma^2-84\gamma+40)}}{2\gamma\sqrt{|3\gamma-1|}}d{\bar t} \nonumber \\
&+\frac{3\sqrt{(2-3\gamma)(2-\gamma)}C_0}{2\sqrt{|3\gamma-1|}}{\bar t}^{2(1-\gamma)/\gamma}d{\bar x},
\end{align}  
where $n_\mu n^\mu={\bar n}_\mu {\bar n}^\mu= +1 (-1)$ for $\gamma>(<)1/3$.
We have set the sign of the normal 1-forms so that they point from the Kantowski-Sachs side to the Friedmann side, i.e., in the directions of increasing $x$ and ${\bar x}$.

The extrinsic curvature of $\Sigma$ is obtained from $K_{ab}:= n_{\mu;\nu}e^\mu_a e^\nu_b$, where $e^\mu_a := \partial x^\mu/\partial y^a$ and $y^a=(\lambda,\theta,\phi)$.
On the Friedmann side, we obtain
\begin{eqnarray}
e^0_ady^a&=&d\lambda,\\
e^1_ady^a&=&-\frac{(2-3\gamma)}{3\gamma}C_1\lambda^{-2/(3\gamma)}d\lambda,\\
e^i_ady^a&=& \delta^i_j dy^j,
\end{eqnarray}
where $i$ and $j$ are indices on $S^2$ and $\gamma_{ij}$ is the unit metric on $S^2$, so $d\Omega^2=\gamma_{ij}dx^idx^j$.
The non-zero components of $K_{ab}$ are 
\begin{eqnarray}
K^\lambda_{~~\lambda}&=&-\frac{2-3\gamma}{3\gamma\sqrt{|3\gamma-1|}}\lambda^{-1},\\
K^i_{~~j}&=&\delta^i_{~~j}\frac{9\gamma^2+6\gamma-4}{6\gamma\sqrt{|3\gamma-1|}}\lambda^{-1}.
\end{eqnarray}
On the Kantowski-Sachs side, we obtain
\begin{eqnarray}
{\bar e}^0_ady^a&=&d\lambda,\\
{\bar e}^1_ady^a&=&-\frac{\sqrt{27\gamma^2-84\gamma+40}}{3\gamma C_0}\lambda^{2(\gamma-1)/\gamma}d\lambda,\\
{\bar e}^i_ady^a&=& \delta^i_j dy^j,
\end{eqnarray}
where ${\bar e}^\mu_a := \partial {\bar x}^\mu/\partial y^a$.
The non-zero components of ${\bar K}_{ab}(:={\bar n}_{\mu;\nu}{\bar e}^\mu_a {\bar e}^\nu_b)$ are 
\begin{eqnarray}
{\bar K}^\lambda_{~~\lambda}&=&-\frac{4(1-\gamma)(3\gamma-1)}{9(2-3\gamma)(2-\gamma)} \nonumber \\
&&\times\sqrt{\frac{27\gamma^2-84\gamma+40}{|3\gamma-1|(2-3\gamma)(2-\gamma)}}\lambda^{-1},\\
{\bar K}^i_{~~j}&=&-\delta^i_{~~j}\frac{\gamma}{2}\sqrt{\frac{27\gamma^2-84\gamma+40}{|3\gamma-1|(2-3\gamma)(2-\gamma)}}\lambda^{-1}.
\end{eqnarray}

It is seen that the second fundamental form $K_{ab}$ is discontinuous at $\Sigma$
in general, while the first fundamental form $h_{ab}$ is continuous.
This means that we can match the two spacetimes with a singular hypersurface 
on $\Sigma$.
As shown below, the matter on the shell has the form of a perfect fluid obeying a linear equation of state.

\begin{center}
\begin{figure}[htbp]
\includegraphics[width=0.5\textwidth]{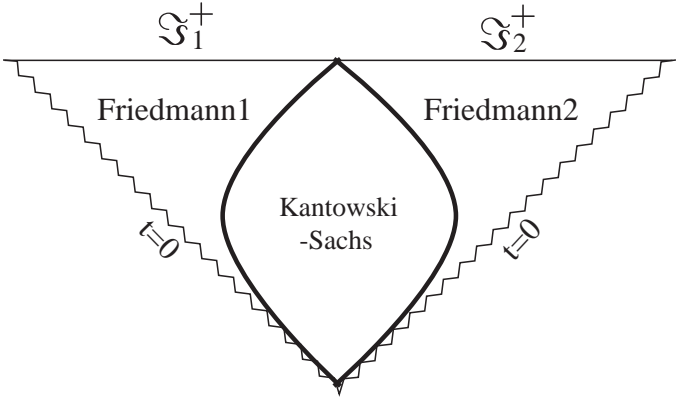}
\caption{\label{fg:F-KS} The causal structure of 
the Friedmann-Friedmann cosmological wormhole spacetime with thin shells for $1/3 < \gamma< (14-2\sqrt{19})/9$.
$t=0$ corresponds to the initial singularity, while 
$\Im^{+}_1$ and $\Im^{+}_2$ correspond to two distinct null infinities.
The thick curves correspond to two massive thin shells with negative 
surface energy density.
The whole spacetime is foliated by trapped surfaces 
and there is no trapping horizon.
}
\end{figure}
\end{center}

The energy-momentum tensor on the shell $S^a_{~~b}$ is 
\begin{eqnarray}
8\pi G S^a_{~~b}=-\varepsilon ([K^a_{~~b}]-[K]h^a_{~~b}),
\end{eqnarray}
where $[X]:=X-{\bar X}$ and $\varepsilon=+1 (-1)$ for $\gamma>(<)1/3$.
As a result, we obtain
\begin{widetext}
\begin{eqnarray}
8\pi G S^\lambda_{~~\lambda}&=&-\varepsilon A \lambda^{-1},\\
A&:=&-\frac{9\gamma^2+6\gamma-4}{3\gamma\sqrt{|3\gamma-1|}}-\gamma\sqrt{\frac{27\gamma^2-84\gamma+40}{|3\gamma-1|(2-3\gamma)(2-\gamma)}}, \\
8\pi G S^i_{~~j}&=&\varepsilon B \lambda^{-1}\delta^i_{~~j},\\
B&:=&\frac{9\gamma^2+12\gamma-8}{6\gamma\sqrt{|3\gamma-1|}}+\frac{27\gamma^3-96\gamma^2+68\gamma-8}{18(2-3\gamma)(2-\gamma)}\sqrt{\frac{27\gamma^2-84\gamma+40}{|3\gamma-1|(2-3\gamma)(2-\gamma)}}.
\end{eqnarray}
\end{widetext}
It can be shown that $A>0$ and $A<0$ for $0<\gamma<1/3$ and $1/3<\gamma<(14-2\sqrt{19})/9$, respectively.
This means that the energy density of the matter on the timelike shell (identified with $-S^\lambda_{~~\lambda}$) is negative and the weak energy condition is violated.
On the other hand, the spacelike matching surface may be regarded as a kind of phase transition.
The Penrose diagram of the resulting spacetime is shown in Fig.~\ref{fg:F-KS} for the case of the timelike shells.

\subsubsection{The null-shell case}
In the $\gamma=1/3$ case, the shell is a null hypersurface and has to be treated separately.
The Friedmann and Kantowski-Sachs metrics are now 
\begin{eqnarray}
ds_{+}^2&=&-dt^2+\biggl(\frac{t}{t_0}\biggl)^{4}(dx^2+x^2d\Omega^2),\\
ds_{-}^2&=&-15d{\bar t}^2+C_0^2{\bar t}^{8}d{\bar x}^2+{\bar t}^2d\Omega^2.
\end{eqnarray}  
The matching null hypersurface is 
\begin{eqnarray}
t=\lambda, \quad 
x=C_1 \lambda^{-1}, \quad \theta=\theta, \quad \phi=\phi
\end{eqnarray}  
on the Friedmann side and 
\begin{eqnarray}
{\bar t}=\lambda,\quad 
{\bar x}=C_2\lambda^{-3}, \quad \theta=\theta, \quad \phi=\phi
\end{eqnarray}  
on the Kantowski-Sachs side.
By continuity of the induced metric, we obtain
\begin{eqnarray}
C_1^2=t_0^{4},\quad 
C_2^2=\frac{5}{3C_0^2}
\end{eqnarray}  
and the induced metric is 
\begin{eqnarray}
ds_{\Sigma}^2&=&\sigma_{ij}dx^i dx^j \\
&=&\lambda^2\gamma_{ij}dx^i dx^j. 
\end{eqnarray}

The radial basis vectors $k^\mu:=\partial x^\mu/\partial \lambda$ and ${\bar k}^\mu:=\partial {\bar x}^\mu/\partial \lambda$ on $\Sigma$ are
\begin{eqnarray}
k^\mu \frac{\partial}{\partial x^\mu}&=&\frac{\partial}{\partial t}-C_1 t^{-2}\frac{\partial}{\partial x},\\
{\bar k}^\mu \frac{\partial}{\partial {\bar x}^\mu}&=&\frac{\partial}{\partial {\bar t}}-3C_2{\bar t}^{-4}\frac{\partial}{\partial {\bar x}},
\end{eqnarray}  
so $k^\mu k_\mu={\bar k}^\mu{\bar k}_\mu=0$ on $\Sigma$.
The basis vectors of $\Sigma$ are 
\begin{align}
e^\mu_\lambda\frac{\partial}{\partial x^\mu}&=k^\mu\frac{\partial}{\partial x^\mu},\\
e^\mu_i\frac{\partial}{\partial x^\mu}&=\delta^\mu_{~~i}\frac{\partial}{\partial x^\mu}
\end{align}
on the Friedmann side and 
\begin{align}
{\bar e}^\mu_\lambda\frac{\partial}{\partial {\bar x}^\mu}&={\bar k}^\mu\frac{\partial}{\partial {\bar x}^\mu},\\
{\bar e}^\mu_i\frac{\partial}{\partial {\bar x}^\mu}&=\delta^\mu_{~~i}\frac{\partial}{\partial {\bar x}^\mu}
\end{align}
on the Kantowski-Sachs side.
The bases are completed by 
\begin{eqnarray}
N_\mu dx^\mu&=&-\frac12 dt+\frac12 C_1^{-1} t^{2}dx,\\
{\bar N}_\mu d{\bar x}^\mu&=&-\frac12d{\bar t}+\frac16 C_2^{-1}{\bar t}^{4}d{\bar x},
\end{eqnarray}  
which satisfy $N^{\mu}N_{\mu}={\bar N}^{\mu}{\bar N}_{\mu}=0$, 
$N_\mu k^\mu={\bar N}_\mu {\bar k}^\mu=-1$ and $N_\mu e^\mu_{i}={\bar N}_\mu {\bar e}^\mu_{i}=0$.

The nonvanishing components of the transverse curvature $C_{ab}:=N_{\mu;\nu} e^\mu_a e^\nu_b$ (${\bar C}_{ab}:={\bar N}_{\mu;\nu} {\bar e}^\mu_a {\bar e}^\nu_b$) are 
\begin{eqnarray}
C_{\lambda\lambda}=\frac{2}{\lambda},\quad 
C_{ij}=\gamma_{ij}\frac{3\lambda}{2}
\end{eqnarray}
in the Friedmann region and 
\begin{eqnarray}
{\bar C}_{\lambda\lambda}=\frac{4}{\lambda},\quad 
{\bar C}_{ij}=\gamma_{ij}\frac{\lambda}{30}
\end{eqnarray}
in the Kantowski-Sachs region.
$C_{\lambda i}=0={\bar C}_{\lambda i}$ means that there is no heat flow 
on the shell.
The pressure and surface energy density of the matter on the shell are given by
\begin{eqnarray}
8\pi G p_{\rm shell}&:=&-[C_{\lambda\lambda}]
=\frac{2}{\lambda},\\
8\pi G \mu_{\rm shell}&:=&-\sigma^{ij}[C_{ij}]
=-\frac{44}{15\lambda},
\end{eqnarray}
respectively. It is seen that the matter on the shell has negative surface 
energy density and violates the weak energy condition, as in the case of a timelike shell.

Although the Friedmann-Friedmann cosmological wormhole numerically obtained in ref.~\cite{mhc1} satisfies the dominant energy condition in the whole spacetime, the matter content in this analytic solution violates the weak energy 
condition on the shell.
This is due to the simplification entailed in assuming a singular hypersurface.
Nevertheless, the solution still provides an analytic example of a cosmological wormhole which is not of the Hochberg-Visser or Hayward type.

\subsection{An analytic solution without a shell}
The analytic solution discussed above contains thin shells.
Next we present an analytic solution
without a thin shell.
The spacetime is asymptotically Friedmann and trapped everywhere, so again it is not a Hochberg-Visser or Hayward wormhole. 

We consider the simple metric
\begin{eqnarray}
ds^2=-dt^2+a(t)^2[dx^2+(x^2+b^2)d\Omega^2], \label{c-ellis}
\end{eqnarray}  
where $b$ is a positive constant.
Such solutions are conformal to the Morris-Thorne wormhole spacetimes studied in refs.~\cite{kim1996,hv1998,cataldo2008,cataldo2009}.
(See also~\cite{snk2008} for the analysis of the wormhole dynamics.)
The spacetime is asymptotically Friedmann for $x \to \pm \infty$ 
with scale factor $a(t)$.
The wormhole throat is located at $x=0$ on a spacelike hypersurface with constant $t$, around which the metric is approximately Kantowski-Sachs.
The corresponding energy-momentum tensor is given by
\begin{align}
8\pi G T^{t}_{~~t}&=-3\frac{{\dot a}^2}{a^2}+\frac{b^2}{a^2(x^2+b^2)^2}=:-8\pi G \mu_{\rm tot},\\
8\pi G T^{x}_{~~x}&=-2\frac{{\ddot a}}{a}-\frac{{\dot a}^2}{a^2}-\frac{b^2}{a^2(x^2+b^2)^2}=:8\pi G p_{r,{\rm tot}},\\
8\pi G T^{\theta}_{~~\theta}&=8\pi G T^{\phi}_{~~\phi} \nonumber \\
&=-2\frac{{\ddot a}}{a}-\frac{{\dot a}^2}{a^2}+\frac{b^2}{a^2(x^2+b^2)^2}=:8\pi G p_{t,{\rm tot}},
\end{align}
where a dot denotes the derivative with respect to $t$.
The matter field is regarded as a mixture of a perfect fluid and a massless ghost scalar field, i.e., a massless scalar field with a negative kinetic term.
The Misner-Sharp mass is given by
\begin{align}
2 G m=a\sqrt{x^2+b^2}\biggl(\frac{b^2}{x^2+b^2}+{\dot a}^2(x^2+b^2)\biggl),
\end{align}
which is positive everywhere.

If $a$ is constant, this spacetime coincides with the static Ellis wormhole~\cite{ellis1973}.
If we also set $a=t/t_0$, where $t_0$ is a positive constant,
there is a null big-bang initial singularity at $t=0$.
The corresponding energy density, radial pressure and the tangential pressure are now given by
\begin{align}
8\pi G \mu_{\rm tot}&=\frac{3}{t^2}-\frac{t_0^2b^2}{t^2(x^2+b^2)^2},\\
8\pi G p_{r,{\rm tot}}&=-\frac{1}{t^2}-\frac{t_0^2b^2}{t^2(x^2+b^2)^2},\\
8\pi G p_{t,{\rm tot}}&=-\frac{1}{t^2}+\frac{t_0^2b^2}{t^2(x^2+b^2)^2},
\end{align}
respectively. The equation of state for the perfect fluid
is therefore $p=-(1/3)\mu$ in this case.
We see that $\mu_{{\rm tot}}-p_{r,{\rm tot}}> 0$ and $\mu_{{\rm tot}}+p_{t,{\rm tot}}> 0$.
Also it can be shown that $\mu_{{\rm tot}}\ge 0$ for $t_0 \le \sqrt{3}b$, $\mu_{{\rm tot}}+p_{r,{\rm tot}}\ge 0$ for $t_0 \le b$, and $\mu_{{\rm tot}}-p_{t,{\rm tot}}\ge 0$ for $t_0 \le \sqrt{2}b$.
Hence, the dominant energy condition is satisfied for $t_0 \le b$.
Because $\mu_{{\rm tot}}+p_{r,{\rm tot}}+2p_{r,{\rm tot}} \equiv 0$, the strong energy condition is also satisfied for $t_0 \le b$.
The trapped condition $r<2Gm$ reduces to 
\begin{align}
(x^2+b^2)^2-t_0^2(x^2+b^2)+b^2t_0^2>0,
\end{align}
which is satisfied everywhere in the spacetime for $t_0 < 2b$.

\begin{center}
\begin{figure}[htbp]
\includegraphics[width=0.4\textwidth]{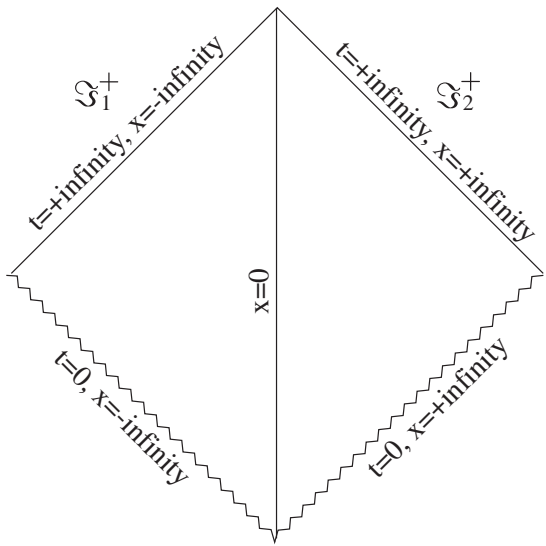}
\caption{\label{fg:ellis} The causal structure of 
the cosmological Ellis wormhole spacetime (\ref{c-ellis}) for $a=t/t_0$.
$t=0$ corresponds to the null initial singularity, while 
$\Im^{+}_1$ and $\Im^{+}_2$ correspond to two distinct null infinities.
The thin straight line corresponds to the wormhole throat $x=0$.
For $t_0< 2b$, the whole spacetime is foliated by trapped surfaces and there is no trapping horizon.
For $t_0 \le b$, the dominant energy condition is satisfied.
}
\end{figure}
\end{center}

In summary, for $t_0< 2b$, the spacetime represents a cosmological wormhole with no trapping horizon.
Moreover, 
for $t_0 \le b$, this cosmological wormhole satisfies the dominant energy condition in the whole spacetime.
The Penrose diagram is shown in Fig.~\ref{fg:ellis}.

Because we have considered the simplest case with $a=t/t_0$, the null infinities are null, so 
the global structure is different from that of the Friedmann-Friedmann cosmological wormhole solution obtained in ref.~\cite{mhc1}.
If we assume $a=(t/t_0)^p$ with $p>1$, corresponding to an accelerating universe, the null infinities are spacelike and the initial singularity remains null.

\section{Summary}
This work is motivated by the cosmological wormhole solutions which we recently obtained numerically~\cite{mhc1,hmc1}.
The dominant energy condition is satisfied in the whole spacetime 
for those solutions and the wormhole throats connect a Friedmann universe at one infinity to another asymptotic solution at the other infinity.
With fine-tuning of the single parameter involved, the wormhole throat connects two Friedmann universes. Nevertheless, the whole spacetime is trapped 
and there is no trapping horizon, so these spacetimes are not Hochberg-Visser or Hayward wormholes.

This has led us to define a wormhole throat on a spacelike hypersurface, since this 
includes our new interesting class of {\it cosmological wormholes}.
We have shown that that dynamical wormhole throat 
may be located in the trapped region.
If the spacetime is asymptotically Friedmann and foliated by trapped surfaces, this implies that  it can contain a wormhole throat with no trapping horizon.
This is impossible in an asymptotically flat dynamical spacetime because the spacetime is foliated by untrapped surfaces near the asymptotically flat region.

We have found an analytic solution corresponding to our numerical Friedmann-Friedmann wormhole.
This is constructed by gluing the Friedmann exterior to the Kantowski-Sachs interior via a massive thin shell under the assumption that the perfect fluid contained in each spacetime obeys the same equation of state $p=(\gamma-1)\mu$.
The dominant energy condition requires $0\le  \gamma \le 2$ and
the Kantowski-Sachs spacetime is Lorentzian for $0< \gamma <2/3$.
The matching is possible for $0<\gamma<(14-2\sqrt{19})/9 \simeq 0.5869 (<2/3)$.
The matching surface $\Sigma$ is timelike for $\gamma>1/3$,  spacelike for $\gamma<1/3$  and null for $\gamma=1/3$. 
The matter on the shell necessarily has a negative energy density for $1/3 \le \gamma< (14-2\sqrt{19})/9$, but the
solution is still interesting because it provides a simple analytic model for a cosmological wormhole which is not in the Hochberg-Visser or Hayward class.

We have also constructed an analytic solution for cosmological wormholes without a massive thin shell.
This solution contains a ghost scalar field and
a perfect fluid. It has a wormhole throat connecting 
two distinct Friedmann universes. With an appropriate 
choice of scale factor, the whole spacetime 
is trapped and the dominant energy condition still holds.

It is found that the 
Kantowski-Sachs dynamical solutions 
are important for these cosmological wormhole spacetimes.
The (quasi-)Kantowski-Sachs solution describes the 
wormhole throat in both our numerical and analytic Friedmann-(quasi-)Friedmann wormhole solutions~\cite{hmc1,mhc1}.
It is conjectured that a cosmological wormhole always has a Kantowski-Sachs structure at the throat.
This class of cosmological wormholes could be important 
in the very early universe.

\acknowledgments

The authors are grateful to S.A.~Hayward for helpful discussion and
useful comments.
HM was supported by Fondecyt grant 1071125.
The Centro de Estudios Cient\'{\i}ficos (CECS) is funded by the Chilean
Government through the Millennium Science Initiative and the Centers of
Excellence Base Financing Program of Conicyt. CECS is also supported by a
group of private companies which at present includes Antofagasta Minerals,
Arauco, Empresas CMPC, Indura, Naviera Ultragas, and Telef\'{o}nica del Sur.
TH was supported by the Grant-in-Aid
for Scientific Research Fund of the Ministry of Education,
Culture, Sports and Technology, Japan (Young Scientists (B) 18740144).
TH was also grateful to CECS for its hospitality during his visit by the Fondecyt grant 7080214.



\begin{thebibliography}{99}
\bibitem{mt1988}
M.S.~Morris and K.S.~Thorne,
Am. J. Phys. {\bf 56}, 395 (1988).
\bibitem{ellis1973}
  H.~G.~Ellis,
  J.\ Math.\ Phys.\  {\bf 14}, 104 (1973).
\bibitem{before}
  H.~G.~Ellis,
  Gen.\ Rel.\ Grav.\  {\bf 10}, 105 (1979);
  K.~A.~Bronnikov,
  Acta Phys.\ Polon.\  B {\bf 4}, 251 (1973);
  T.~Kodama,
  Phys.\ Rev.\  D {\bf 18}, 3529 (1978);
  G.~Cl{\'e}ment,
  Gen.\ Rel.\ Grav.\  {\bf 13}, 763 (1981).
\bibitem{visser}
M. Visser, {\it Lorentzian Wormholes: From Einstein to Hawking},
(Springer-Verlag, Berlin, Germany, 1997).
\bibitem{superluminal}
  M.~Visser, B.~Bassett and S.~Liberati,
  arXiv:gr-qc/9908023;
  M.~Visser, B.~Bassett and S.~Liberati,
  Nucl.\ Phys.\ Proc.\ Suppl.\  {\bf 88}, 267 (2000);
\bibitem{lobo2007}
F.S.N.~Lobo, 
e-Print: arXiv:0710.4474 [gr-qc]. 
\bibitem{mty1988}
M.S.~Morris, K.S.~Thorne, and U.~Yurtsever, 
Phys. Rev. Lett. {\bf 61}, 1446 (1988).
\bibitem{timemachine}
M.~Visser,
Phys. Rev. D{\bf 47}, 554 (1993);
S.W.~Kim and K.S.~Thorne, 
Phys. Rev. D{\bf 43}, 3929 (1991). 
\bibitem{hv1997}
D.~Hochberg and M.~Visser,
Phys. Rev. D{\bf 56}, 4745 (1997).
\bibitem{negative}
D.~Ida and S.A.~Hayward, 
Phys.Lett. {\bf A260}, 175 (1999); 
M.~Visser, S.~Kar, and N.~Dadhich,
Phys. Rev. Lett. {\bf 90}, 201102 (2003);
C.J.~Fewster and T.A.~Roman,
Phys. Rev. D{\bf 72}, 044023 (2005); 
P.K.F.~Kuhfittig,
Phys. Rev. D{\bf 73}, 084014 (2006); 
O.B.~Zaslavskii, 
Phys. Rev. D{\bf 76}, 044017 (2007). 
\bibitem{TC}
  J.~L.~Friedman, K.~Schleich and D.~M.~Witt,
  Phys.\ Rev.\ Lett.\  {\bf 71}, 1486 (1993)
  [Erratum-ibid.\  {\bf 75}, 1872 (1995)];
%
  G.~J.~Galloway, K.~Schleich, D.~M.~Witt and E.~Woolgar,
  Phys.\ Rev.\  D {\bf 60}, 104039 (1999).
\bibitem{vkd2003}
M.~Visser, S.~Kar, and N.~Dadhich,
Phys. Rev. Lett. {\bf 90}, 201102 (2003).
\bibitem{hv1998}
D.~Hochberg and M.~Visser,
Phys. Rev. D{\bf 58}, 044021 (1998).
\bibitem{hayward1999}
S.A.~Hayward,
Int. J. Mod. Phys. D{\bf 8}, 373 (1999).
\bibitem{hayward1994}
S.A.~Hayward,
Phys. Rev. D{\bf 49}, 6467 (1994).
\bibitem{hayward}
S.A.~Hayward,
private communication.
\bibitem{mhc1}
H.~Maeda, T.~Harada and B.J.~Carr, 
Phys. Rev. D{\bf 77}, 024023 (2008).
\bibitem{hmc1}
T.~Harada, H.~Maeda and B.J.~Carr, 
Phys. Rev. D{\bf 77}, 024022 (2008).
\bibitem{hayward1996}
S.A.~Hayward,
Phys. Rev. D{\bf 53}, 1938 (1996).
\bibitem{ms1964} 
C.W.~Misner and D.H.~Sharp, 
Phys. Rev. {\bf 136}, B571 (1964).
\bibitem{Poisson}
E.~Poisson, 
{\it A Relativist's Toolkit}
(Cambridge University Press, Cambridge, England, 2004).
\bibitem{kim1996}
S.-W.~Kim, 
Phys. Rev. D{\bf 53}, 6889 (1996).
\bibitem{cataldo2008}
M.~Cataldo, P.~Labrana, S.~del Campo, J.~Crisostomo, and P.~Salgado,
Phys. Rev. D{\bf 78}, 104006 (2008).
\bibitem{cataldo2009}
M.~Cataldo, S.~del Campo, P.~Minning, and P.~Salgado,
e-Print: arXiv:0812.4436 [gr-qc] 
\bibitem{snk2008}
A.~Shatskiy, I.D.~Novikov, and N.S.~Kardashev,
Phys. Usp. {\bf 51}, 457 (2008), 
e-Print: arXiv:0810.0468 [gr-qc].

\end{thebibliography}
\end{document}